\documentclass[12pt]{article}
\usepackage{a4}
\usepackage{amsfonts,amssymb,amsmath}
\usepackage{graphicx}
\usepackage{cite}

\begin{document}

\title{{\Large Dirac equation on coordinate dependent noncommutative space-time}}
\author{V.G. Kupriyanov\thanks{%
e-mail: vladislav.kupriyanov@gmail.com} \\
CMCC, Universidade Federal do ABC, Santo Andr\'{e}, SP, Brazil}
\date{\today                                        }
\maketitle

\begin{abstract}
In this paper we discuss classical aspects of spinor field theory on the coordinate dependent noncommutative space-time. The noncommutative Dirac equation describing spinning particle in an external vector field and the corresponding action principle are proposed. The specific choice of a star product allows us to derive a conserved noncommutative probability current and to obtain the energy-momentum tensor for free noncommutative spinor field. Finally, we consider a free noncommutative Dirac fermion and show that if the Poisson structure is Lorentz-covariant, the standard energy-momentum dispersion relation remains valid.
\end{abstract}

\section{Introduction}

The idea to use noncommutative (NC) coordinates in quantum mechanics
appeared a long time ago. Peierls used noncommutative coordinates in \cite{Peierls} to describe a charged particle in strong magnetic field and in the
presence of a weak electric potential. In \cite{Snyder} Snyder proposed an example of a Lorentz invariant noncommutative space-time. Since the commutator between the coordinates in \cite{Snyder} is proportional to the generator of Lorentz transformation, the coordinate operators do not form a subalgebra. So, if the field is understood as a function of coordinates, then the product of two fields will not be a field anymore, since it will not depend only on coordinates, but also on momenta. Later the noncommutativity was suggested as a mathematical tool for regularization.

In last decades, motivated by the string theory \cite{SW} and quantum gravity \cite{Freidel}, the subject gained a lot of interest and
has been studied extensively \cite{NCreviews}. The main attention, however, was given to
the case of canonical noncommutative space-time, realized by coordinate
operators $\hat{x}^{\rho},\ \rho=0,...,N-1,$ satisfying the algebra $\left[ \hat{x}%
^{\rho},\hat {x}^{\sigma}\right] =i\theta^{\rho\sigma}\,,$ with a constant $\theta^{\rho\sigma}$
being the parameters of noncommutativity. Various phenomenological
consequences of the presence of this type of noncommutativity in the theory
were studied.

One of the main problems of the canonical noncommutativity is the violation of physical symmetries, like the Lorentz symmetry. Different approaches to the solution of this problem may be found in the literature, like twisted symmetries \cite{Chaichian}. However the discussion about physical consequences of the twisted symmetries is open, see \cite{Pinzul}. For some other examples of Lorentz invariant noncommutative theories see \cite{NCLorentz} and references therein.

Physical
motivations of the noncommutativity come from a combination of
quantum mechanics with Einstein relativity \cite{Doplicher}. While
space-time in any dynamical theory of gravity cannot be flat. So, the restriction
to the canonical noncommutative spaces does not seem to be very natural.
The presence of a more general type of NC spaces with the
parameters of noncommutativity depending on coordinates may lead to absolutely
different phenomenological consequences. For the particular examples of coordinate dependent noncommutativity in the field theory see the following papers and references therein. In \cite{Steinacker} the noncommutativity appears in the context of matrix models. A field theoretical models on kappa-Minkowski space-time were discussed in \cite{Meljanac}, a quantum field theory on fuzzy spaces was studied in \cite{Vitale}.

In general case we may suppose that the operators of coordinate $\hat{x}^{\rho}$ satisfy the algebra
\begin{equation}
\left[ \hat{x}^{\rho},\hat{x}^{\sigma}\right] =i\theta\hat{\omega}_{q}^{\rho\sigma}\left(
\hat{x}\right) ,  \label{1}
\end{equation}
where $\hat{\omega}_{q}^{\rho\sigma}\left( \hat{x}\right) $ is an operator defined
from physical considerations and the parameter of noncommutativity is given by $\theta= l^2,$ with $l$ being a short-distance length scale. Saying this, we mean that the antisymmetric (dimensionless) field $\omega^{\rho\sigma}\left( x\right)$
obeying the Jacobi identity%
\begin{equation}
\omega^{\rho\nu}\partial_{\nu}\omega^{\sigma\lambda}+\omega^{\lambda\nu}\partial_{\nu}\omega^{\rho\sigma}+%
\omega^{\sigma\nu}\partial_{\nu}\omega^{\lambda \rho}=0,  \label{2}
\end{equation}
which corresponds to the symbol of the operator $\hat{\omega}_{q}^{\rho\sigma}\left( \hat{x}\right) $ is given, and the ordering of the operator $\hat{\omega}_{q}$ is specified.

The aim of this work is to propose a consistent deformation of the relativistic quantum mechanics introducing noncommutativity of the general form (\ref{1}). In Sec. 2 we describe the nonrelativistic quantum mechanics with coordinate dependent noncommutativity (\ref{1}) and investigate some basic properties of the star product. In Sec. 3, we construct a relativistic generalization of this model, introducing a noncommutative Dirac equation and deriving the corresponding action principle. In Sec. 4 we propose a method to derive conserved quantities on coordinate dependent noncommutative spaces. This method is applied to obtain the continuity equation for the noncommutative probability current. Then we derive an expression for the corresponding energy-momentum tensor. Finally, in Sec. 5 we study as an example free noncommutative relativistic spinning particle.

\section{Quantum mechanics with coordinate dependent noncommutativity}

A nonrelativistic version of quantum mechanics on coordinate dependent noncommutative spaces was recently proposed in \cite{kup14,kup15} and can be defined as follows. Suppose that there exists a
function $\mu \left( x\right)\neq 0 $ such that
\begin{equation}
\partial _{\rho}\left( \mu \omega ^{\rho\sigma}\right) =0.  \label{3}
\end{equation}%
Note that if $\det\omega^{\rho\sigma}\neq0$,
the function $\mu \left( x\right) $ can be chosen as
$\mu \left( x\right) =\left\vert \det\omega^{\rho\sigma}(x)\right\vert^{-1/2}$. If the external antisymmetric field $\omega^{\rho\sigma}$ is degenerate or has a nonconstant rank, one should solve the equation (\ref{3}) to obtain the expression for $\mu \left( x\right) $.

Let us define  the Hilbert space as a space of
complex-valued functions which are square-integrable with a measure
\begin{equation} \mathbf{\Omega }\left( x\right) =d^{N}x\mu \left( x\right) .\label{measure}
\end{equation}
The internal product between two states $%
\varphi \left( x\right) $ and $\psi \left( x\right) $ from the Hilbert space
is determined by the formula
\begin{equation}
\left\langle \varphi \right\vert \left. \psi \right\rangle =\int\mathbf{%
\Omega }\left( x\right) \left( \varphi
^{\ast }\star \psi  \right),  \label{scalar}
\end{equation}%
where $\star$ is a star product. For our purposes we choose a specific star product, that should be closed with respect to the measure (\ref{measure}), see \cite{kup15} for such a construction. For any two functions $f$ and $g$ it can be taken in the form:
\begin{align}
& (f\star g)(x)=f\cdot g+\frac{i\theta }{2}\partial
_{\rho}f\omega ^{\rho \sigma}\partial _{\sigma}g -\frac{\theta ^{2}}{8}\omega ^{\rho \sigma}\omega ^{\alpha \beta}\partial
_{\rho}\partial _{\alpha}f\partial _{\sigma}\partial _{\beta}g \label{star} \\
& -\frac{\theta^2}{12}\omega
^{\rho \sigma}\partial _{\sigma}\omega ^{\alpha \beta}\left( \partial _{\rho}\partial _{\alpha}f\partial
_{\beta}g-\partial _{\alpha}f\partial _{\rho}\partial _{\beta}g\right) -\frac{\theta ^{2}}{24\mu}\partial _{\alpha}\left( \mu \omega ^{\rho \sigma}\partial
_{\sigma}\omega ^{\alpha\beta}\right)\partial _{\rho}f\partial
_{\beta}g +O\left(
\theta ^{3}\right) .  \notag
\end{align}
The action of the coordinate operators $\hat{x}^{\rho}$ on functions $\psi (x)$ from
the Hilbert space is defined through the star product (\ref{star}), for any function $%
V\left( x\right) $ one has
\begin{equation}
{V}\left( \hat{x}\right) \psi (x)=V(x)\star \psi (x).  \label{15}
\end{equation}%
In particular, from $\hat x^\rho \psi=x^\rho\star\psi$, one may see that
\begin{equation}
\hat x^\rho=x^\rho+\frac{i\theta}{2}\omega^{\rho \sigma}\partial_\sigma+\frac{\theta^2}{12}\omega^{\alpha\sigma}\partial_\sigma\omega^{\rho \beta}\partial_\alpha\partial_\beta-\frac{\theta^ 2}{24\mu}\partial _{\beta}\left( \mu \omega ^{\rho \sigma}\partial
_{\sigma}\omega ^{\beta\alpha}\right)\partial_\alpha+O\left( \theta ^{3}\right).  \label{15a}
\end{equation}
The definitions (\ref{scalar}) and (\ref{15}) means that the coordinate
operators are self-adjoint with respect to the introduced scalar product: $%
\left\langle \hat{x}^{\rho}\varphi \right\vert \left. \psi \right\rangle
=\left\langle \varphi \right\vert \left. \hat{x}^{\rho}\psi \right\rangle $.
The momentum operators $\hat{p}_{\rho}$ are fixed from the condition
that they also should be self-adjoint with respect to (\ref{scalar}). Choosing it in the form
\begin{equation}
\hat{p}_{\rho}=-i\partial _{\rho}-\frac{i}{2}\partial _{\rho}\ln \mu \left( x\right) ,
\label{p}
\end{equation}%
we obtain
\begin{eqnarray}
&& \left\langle \hat{p}_{\rho}\varphi
\right\vert \left. \psi \right\rangle =\int\mathbf{%
\Omega }\left( x\right) \left[\left(\hat{p}_{\rho} \varphi\right)
^{\ast }\star \psi  \right]=\int\mathbf{%
\Omega }\left( x\right) \left[\left(\hat{p}_{\rho} \varphi\right)
^{\ast }\cdot \psi  \right]=\label{pp}\\
&&\int\mathbf{%
\Omega }\left( x\right) \left[ \varphi
^{\ast }\cdot\left( \hat{p}_{\rho}\psi \right) \right]=\int\mathbf{%
\Omega }\left( x\right) \left[ \varphi
^{\ast }\star \left( \hat{p}_{\rho}\psi \right) \right]=\left\langle \varphi \right\vert
\left. \hat{p}_{\rho}\psi \right\rangle .\nonumber
\end{eqnarray}

The momentum operators (\ref{p}) commute, $[\hat{p}_{\rho},\hat{p}_{\sigma}]=0$. The commutator between $\hat{x}^{\rho}$, defined in (\ref{15a}), and $\hat{p}_{\sigma}$ is
\begin{equation}
\left[ \hat{x}^{\rho},\hat{p}_{\sigma}\right] =i\delta^\rho_\sigma-\frac{i\theta}{2}\left( \partial_\sigma\omega^{\rho \alpha}\left(\hat{x}%
\right)\hat{p}_{\alpha}+ \frac{i}{2}\partial_\sigma \left(\omega^{\rho \alpha}\partial_\alpha\ln \mu\right) \left( \hat x\right) \right)+O\left( \theta ^{2}\right).  \label{xp}
\end{equation}
So, the complete algebra of commutation relations involving $\hat{x}^{\rho}$ and $\hat{p}_{\sigma}$ is a deformation in $\theta$ of a standard Heisenberg algebra.

Let us list here some important properties of the star product (\ref{star}). First of all, it is an associative product and satisfies the relation
\begin{equation}\label{ast}
  (f\star g)^\ast=g^\ast\star f^\ast,
\end{equation}
here $\ast$ stands for the complex conjugation. As it was already mentioned, due to its definition \cite{kup15}, the star product (\ref{star}) is closed, i.e.,
\begin{equation}\label{trace}
  \int\mathbf{%
\Omega }\left( x\right) \left( f\star g\right)=\int\mathbf{%
\Omega }\left( x\right)  f\cdot g.
\end{equation}
It means that
\begin{equation}\label{trace1}
  \mu\cdot \left(f\star g\right)=\mu\cdot  f\cdot g+\partial_\rho a^\rho(f,g),
\end{equation}
where
\begin{align}
& a^\rho(f,g)=\frac{i\theta\mu}{4}\omega^{\rho\sigma}( f\partial_\sigma g-\partial_\sigma f g)+\frac{\theta^2\mu}{16}\omega ^{\rho \sigma}\omega ^{\alpha \beta}\left(\partial_\sigma\partial_\alpha f\partial_\beta g-\partial_\alpha f\partial_\sigma\partial_\beta g\right)\label{trace2}\\
&+\frac{\theta^2\mu}{48}\left(\omega^{\beta\sigma}\partial_\sigma\omega^{\alpha\rho}-\omega^{\alpha\sigma}\partial_\sigma\omega^{\rho\beta}\right)\partial_\alpha f\partial_\beta g+O\left( \theta ^{3}\right).\notag
\end{align}
Now differentiating the both sides of (\ref{trace1}) one finds:
\begin{equation}
 \partial_\alpha[ \mu \cdot(f\star g)]=i\mu\cdot (\hat p_\alpha f)\cdot g+i\mu\cdot f\cdot (\hat p_\alpha  g)+\partial_\rho \partial_\alpha a^\rho(f,g).\label{p1}
\end{equation}
Using (\ref{trace1}) one more time we end up with
\begin{equation}
 i\mu\cdot[ (\hat p_\alpha f)\star g]+i\mu\cdot[ f\star (\hat p_\alpha  g)]=\partial_\alpha[ \mu\cdot( f\star g)]+\partial_\rho b^\rho_\alpha (f,g),  \label{p2}
\end{equation}
where
\begin{align}
&b^\rho_\alpha(f,g)=ia^\rho\left(\hat p_\alpha f,g\right)+ia^\rho\left( f,\hat p_\alpha g\right)-\partial_\alpha a^\rho(f,g)\label{b}\\&=\frac{i\theta\mu}{4}\left(\partial_\alpha\omega^{\rho\sigma}-\partial_\alpha(\ln\mu)\omega^{\rho\sigma}\right)( \partial_\sigma f\cdot g -f\cdot\partial_\sigma g)+O\left( \theta ^{2}\right).\notag
\end{align}
As we will see, the property (\ref{p2}) of the star product (\ref{star}) is crucial for the derivation of the conservation laws in our approach.

Finally, note that all constructions in the present approach are formal perturbative series and we do not discuss the convergence here. To study non-perturbative consequences of the underlying noncommutative space-time one will need an explicit form of the corresponding star product, like in \cite{Presnajder}.

\section{Relativistic generalization}

Now let us suppose that the antisymmetric field $\omega^{\rho\sigma}(x)$ is a dynamical field, describing new quantum degrees of freedom like, e.g., in \cite{Amorim,Verlinde}. In this case, it transforms as a two tensor with respect to a Lorentz group, i.e.,
\begin{equation}
\omega^{\rho\sigma}(x)\rightarrow\omega'^{\rho\sigma}(x')=\Lambda_{\alpha}^{\rho}\omega^{\alpha\beta}(x)\Lambda_{\beta}^{\sigma}.\label{ltomega}
 \end{equation}
So, the operators $\hat{x}^{\rho }$ and $\hat{p}_{\rho } $, defined in (\ref{15a}) and (\ref{p}) will transform as vectors, \begin{equation}
\hat{x}^{\rho }\rightarrow\hat{x}'^{\rho }=\Lambda^{\rho}_{\alpha}\hat{x}^{\alpha },\,\,\,\hat{p}_{\rho }\rightarrow\hat{p}'_{\rho }=\Lambda^{\alpha}_{\rho}\hat{p}_{\alpha}.\label{ltxp}
  \end{equation}
Using this fact we may define the relativistic wave equations on coordinate dependent noncommutative space-time.

Let us consider the spinning particle in an external electromagnetic field defined by the vector potential $A^\rho(x)$ on the coordinate dependent noncommutative space. We postulate the corresponding Dirac equation as follows:
\begin{equation}
-\gamma^\rho \hat {p}_\rho\psi+e\gamma^\rho A_\rho\star\psi -m\psi=0,\label{Dirac}
\end{equation}
where $\gamma^\rho$ are the Dirac matrices satisfying $\left\{\gamma^\rho,\gamma^\sigma\right\}=2\eta^{\rho\sigma}.$
One may check that this equation is covariant under the Lorentz transformations (\ref{ltomega}), (\ref{ltxp}) and%
\begin{equation}
\psi
\rightarrow\psi^{\prime}\left(  x^{\prime}\right)  =S\left(  \Lambda\right)
\psi\left(  x\right)  ,\ \ A^{\rho}\rightarrow A^{\prime\rho}\left(  x^{\prime
}\right)  =\Lambda_{\sigma}^{\rho}A^{\sigma}\left(  x\right)  ,
\end{equation}
where $S\left(  \Lambda\right)  $ belongs to the usual spinor representation
of the Lorentz group.

Action principle yielding the equation (\ref{Dirac}) can be written as
\begin{equation}
S= \int\mathbf{\Omega }\left( x\right)\left[-\frac{1}{2}\bar\psi\star\gamma^\rho \hat {p}_\rho\psi+\frac{1}{2}\hat {p}_\rho\bar\psi\gamma^\rho\star\psi+e\bar\psi\star\gamma^\rho A_\rho\star\psi -m\bar\psi\star\psi\right].\label{Action}
\end{equation}
Taking into account the properties of the star product (\ref{star}) and the operator of momenta $\hat {p}_\rho$ discussed in the previous section one may see that this action is real. Using the equations (\ref{pp}), (\ref{trace}) and the associativity of the star product the action (\ref{Action}) can be represented as
 \begin{equation}
S= \int\mathbf{\Omega }\left( x\right)\bar\psi\left[-\gamma^\rho \hat {p}_\rho\psi+e\gamma^\rho A_\rho\star\psi -m\psi\right].\label{Action1}
\end{equation}
The variation of this action w.r.t. the conjugate spinor $\bar\psi$ gives the noncommutative Dirac equation (\ref{Dirac}). Similar considerations show that the variation of the (\ref{Action}) w.r.t. the spinor $\psi$ lead to the equation
\begin{equation}
\hat p_\rho\bar\psi\gamma^\rho+e\bar\psi\star\gamma^\rho A_\rho -m\bar\psi=0,\label{DiracCj}
\end{equation}
which is a conjugate Dirac equation.

\section{Conservation laws}

On the classical level the action (\ref{Action}) can be used to study the global and local symmetries of the system and to derive the corresponding Noether currents. However, due to the presence of higher derivatives in the corresponding Lagrangian density the direct application of the Noether theorem may be quite tedious problem. As we will see below, sometimes it is easier to use alternative techniques to obtain the conserved quantities.

The main idea of our approach is to derive the classical conservation laws for noncommutative theories with non-constant Poisson structure using the corresponding equations of motion, the properties of the star product (\ref{star}) and the requirement that in the commutative limit, $\theta\rightarrow0$ and $\mu=1$, the conserved noncommutative current should reproduce corresponding commutative physical quantity.

\subsection{Continuity equation}

According to the above prescription we will look for the noncommutative probability current $j^{\rho}_\theta(x)$ in the form $j^{\rho}_\theta=j^{\rho}_0+O(\theta)$, where $j^{\rho}_0=e\bar\psi_0\gamma^\rho\psi_0$ is a standard probability current for the fermionic field. The total derivative of the commutative electric current $e\partial_\rho (\bar\psi_0\gamma^\rho\psi_0)$ can be written as
\begin{eqnarray}
 &&e\partial_\rho \bar\psi_0\gamma^\rho\psi_0+iem\bar\psi_0\psi_0-iem\bar\psi_0\psi_0\notag\\
 &&+ie^2\bar\psi_0\gamma^\rho A_\rho\psi-ie^2\bar\psi_0\gamma^\rho A_\rho\psi+e\bar\psi_0\gamma^\rho\partial_\rho\psi_0. \notag
\end{eqnarray}
The above expression in turn can be represented as
\begin{eqnarray}
&&-i e \left(i\partial_\rho\bar\psi_0\gamma^\rho+e\bar\psi_0\gamma^\rho A_\rho - m\bar\psi_0\right)\psi_0 \label{id0}\\
&&+i e\bar\psi_0\left(-i\gamma^\rho \partial_\rho\psi_0+e\gamma^\rho A_\rho\psi_0 -m\psi_0 \right)=0. \notag
\end{eqnarray}
This proves that the total derivative of the commutative probability current $j^{\rho}_0$ vanishes on-shell. That is, in the commutative case the continuity equation $\partial_\rho j^{\rho}_0=0$ can be obtained from a combination (\ref{id0}) of the Dirac equation and its conjugate.

To derive the conserved noncommutative current we will use the same logic. Let us substitute in the eq. (\ref{id0}) the standard Dirac equation and its conjugate by the noncommutative ones, the pointwise multiplication by the star multiplication and also multiply the resulting identity by the function $\mu(x)$ to be able to use the eq. (\ref{p2}). We end up with
\begin{eqnarray}
&& -ie\mu \bar\psi\star\left(-\gamma^\rho \hat {p}_\rho\psi+e\gamma^\rho A_\rho\star\psi -m\psi \right)\\ \label{id1}
&& +ie\mu \left(\hat p_\rho\bar\psi\gamma^\rho+e\bar\psi\star\gamma^\rho A_\rho -m\bar\psi\right)\star\psi=0. \notag
\end{eqnarray}
Now taking into account the associativity of the star product (\ref{star}), the above equation becomes
\begin{equation}
ie\mu\bar\psi\star\gamma^\rho\hat p_\rho\psi+ie\mu\hat p_\rho\bar\psi\gamma^\rho\star\psi=0.\label{id3}
\end{equation}
Then, the property (\ref{p2}) implies that the left-hand side of the equation (\ref{id3}) can be written as:
\begin{equation}
ie\mu\bar\psi\star\gamma^\rho\hat p_\rho\psi+ie\mu\hat p_\rho\bar\psi\gamma^\rho\star\psi=\partial_\rho j^{\rho}_\theta,\label{id4}
\end{equation}
where
\begin{equation}
j^{\rho}_\theta=e\mu\bar\psi\gamma^\rho\star\psi+eb^\rho_\nu(\bar\psi\gamma^\nu,\psi).\label{current}
\end{equation}
To obtain (\ref{current}) we have also used the fact that gamma matrices do not depend on $x$, i.e., one may write $\bar\psi\gamma^\rho\star\psi=\bar\psi\star\gamma^\rho\psi$ and $b^\rho_\nu(\bar\psi\gamma^\nu,\psi)=b^\rho_\nu(\bar\psi,\gamma^\nu\psi).$ Taking into account the identity (\ref{id4}) the equation (\ref{id3}) becomes
\begin{equation}
\partial_\rho j^{\rho}_\theta=0.\label{ce}
\end{equation}
By the construction in the commutative limit the vector $j^{\rho}_\theta$ tends to the standard current density for the commutative spinor field, $j^{\rho}_0$. According to the equation (\ref{ce}) this quantity is conserved. So, we call it as the noncommutative current density. The equation (\ref{ce}) is the continuity equation. See \cite{Nowak} for the definition of the conserved currents in the field theory on kappa-Minkowski space. In \cite{Matrix} the conserved currents were determined in matrix models.

 The probability density is determined as zero component of the noncommutative current density,
 \begin{equation}
\varrho(x)=j^{0}_\theta=e\mu\psi^\dagger\star\psi+eb^0_\nu(\bar\psi\gamma^\nu,\psi).\label{density}
\end{equation}
Note that because of the continuity equation (\ref{ce}) the integral over the classical hypersurface \begin{equation}
\int\!\varrho(x)\,d^{N-1}x,\notag\end{equation} which determines the electric charge for the given configuration of fields is a constant in time. That is, as it was expected, the electric charge in the present system is conserved.

\subsection{Energy-momentum tensor}

To obtain an expression for the noncommutative energy-momentum tensor one may apply the same technique as above. For the free commutative fermionic field the stress-energy tensor is known to be
\begin{equation}
T^{\rho\nu}_0=i\bar\psi\gamma^\rho\partial^\nu\psi-i\delta^{\rho\nu}\bar\psi\gamma^\sigma\partial_\sigma\psi+\delta^{\rho\nu} m \bar\psi\psi.\label{T0}
\end{equation}
Its total derivative can be written as:
\begin{equation}
 \partial^\nu\bar\psi\left(-i\gamma^\rho \partial_\rho\psi +m\psi \right)+ \left(\partial_\rho\bar\psi\gamma^\rho +m\bar\psi\right)\partial^\nu\psi=0. \label{id50}
\end{equation}
It means that the conservation law of the free energy-momentum tensor, $\partial_\rho T^{\rho\nu}_0=0,$ is a consequence of the combination (\ref{id50}) of the free commutative Dirac equation and its conjugate.

Substituting in the equation (\ref{id50}) the free Dirac equation and its conjugate by the corresponding noncommutative analogues, using the star product instead of standard product and multiplying the both sides of the identity by the function $\mu(x)$ one obtains:
\begin{equation}
-i\mu \hat {p}^\nu\bar\psi\star\left(\gamma^\rho \hat {p}_\rho\psi +m\psi \right)+i\mu \left(\hat p_\rho\bar\psi\gamma^\rho -m\bar\psi\right)\star\hat {p}^\nu\psi=0. \label{1d5}
\end{equation}
Let us represent (\ref{1d5}) in the form
\begin{eqnarray}
&&-\frac{1}{2}\left(i\mu \hat {p}^\nu\bar\psi\star\gamma^\rho \hat {p}_\rho\psi+i\mu\bar\psi\star\hat {p}^\nu\gamma^\rho \hat {p}_\rho\psi\right)\label{id6} \\
&& +\frac{1}{2}\left(i\mu\hat {p}_\rho\bar\psi\gamma^\rho\star\hat {p}^\nu\psi +i\mu\bar\psi\star\hat {p}_\rho\gamma^\rho\hat {p}^\nu\psi\right) \notag\\
&&+\frac{1}{2}\left(i\mu\hat {p}^\nu\hat p_\rho\bar\psi \gamma^\rho\star\psi +i\mu \hat p_\rho\bar\psi\gamma^\rho \star\hat {p}^\nu\psi\right) \notag\\
&&-\frac{1}{2}\left(i\mu\hat p_\rho\hat {p}^\nu\bar\psi \gamma^\rho\star\psi+i\mu \hat {p}^\nu\bar\psi\star\gamma^\rho \hat {p}_\rho\psi\right) \notag\\
&& +im\mu \hat {p}^\nu\bar\psi\star\psi+im\mu \bar\psi\star\hat {p}^\nu\psi=0.\notag
\end{eqnarray}
Now using the property (\ref{p2}) we represent the left-hand side of (\ref{id6}) as a total derivative:
\begin{equation}\label{T}
\partial_\rho T^{\rho\nu}_\theta=0,
\end{equation}
where
\begin{eqnarray}
&& T^{\rho\nu}_\theta= \frac{\mu}{2}\left(\bar\psi\gamma^\rho\star\hat {p}^\nu\psi-\hat {p}^\nu\bar\psi\star\gamma^\rho \psi\right) \label{id7}
\\&& -\frac{\mu}{2}\delta^{\rho\nu}\left( \bar\psi\star\gamma^\beta \hat {p}_\beta\psi-\hat p_\beta\bar\psi\gamma^\beta \star\psi-2m \bar\psi\star\psi\right)\notag\\&& +\frac{1}{2}b^\rho_\beta\left(\bar\psi,\gamma^\beta\hat{p}^\nu\psi\right)-\frac{1}{2}b^\rho_\beta\left(\hat{p}^\nu\bar\psi\gamma^\beta,\psi\right) \notag
\\&&
-\frac{1}{2}b^{\rho\nu}\left(\bar\psi,\gamma^\beta\hat{p}_\beta\psi\right)+\frac{1}{2}b^{\rho\nu}\left(\hat{p}_\beta\bar\psi\gamma^\beta,\psi\right)
+mb^{\rho\nu}\left(\bar\psi,\psi\right), \notag
\end{eqnarray}
Again, in the commutative limit by the construction, the tensor $T^{\rho\nu}_\theta$ tends to the standard energy-momentum tensor $T^{\rho\nu}_0$ for the commutative spinor field. The equation (\ref{T}) means that $T^{\rho\nu}_\theta$ is a conserved quantity. So, we call it as the energy-momentum tensor for the free noncommutative spinor field.

As usual we define the energy and the momentum of the system by integrating their densities, $T^{00}_\theta$ and $T^{0i}_\theta$, over the corresponding hypersurfaces
\begin{equation}\label{D}
  E=\int\! T^{00}_\theta(x)\,d^{N-1}x,\,\,\,\,\,P_i=\int\! T^{0i}_\theta(x)\,d^{N-1}x.
\end{equation}
Due to the equation (\ref{T}) these quantities are conserved.

In conclusion we stress that the above analysis and results are valid for the noncommutative theories corresponding to any Poisson structure, either Lorentz covariant or not.

\section{Free noncommutative fermionic field}

In this section we study the solution of the free noncommutative Dirac equation
\begin{equation}
\left[i \gamma^\rho\left(\partial_\rho+\frac{1}{2}\partial_\rho\ln\mu(x)\right)-m\right]\psi(x)=0.\label{free}
\end{equation}
It is important to note that if $\psi_0(x)$ is a solution of a free commutative Dirac equation $[i \gamma^\rho\partial_\rho-m]\psi_0(x)=0$, then
\begin{equation}
\psi(x)=\frac{1}{\sqrt{\mu(x)}}\psi_0(x)\label{freesolution}
\end{equation}
is a solution of the free noncommutative Dirac equation (\ref{free}). However, due to the modification of the expression for the probability density (\ref{density}) in the noncommutative case, the normalization factor $N$ should change.

Consider a noncommutative spinning particle in $(2+1)$ dimensions. In this case the antisymmetric field which determines noncommutativity can be chosen as
\begin{equation}\label{omega}
    \omega ^{\rho \sigma}(x)=\varepsilon^{\rho \sigma\tau}x_\tau.
\end{equation}
Since $\partial_\rho\omega ^{\rho \sigma}(x)=\partial_\rho\varepsilon^{\rho \sigma\tau}x_\tau=\varepsilon^{\rho \sigma}_\rho=0$, the function $\mu(x)$ from the definition of measure (\ref{measure}) can be chosen as a a constant, so we set $\mu(x)=1$. That is, $\hat{p}_{\rho}=-i\partial _{\rho}$. Dirac gamma matrices in $(2+1)$ dimensions can be represented by the Pauli sigma matrices. Let $\gamma^0=\sigma_z$, $\gamma^1=i\sigma_x$ and $\gamma^2=i\sigma_y$. In these notations the solution of the free noncommutative Dirac equation reads
\begin{equation}
\psi(x)=N\left(
\begin{array}{c}
  1 \\
  \frac{ip_1-p_2}{\epsilon+m}
\end{array}
\right)e^{-ip_\rho x^\rho},\label{s}
\end{equation}
where $p_\rho=(\epsilon,-p_1,-p_2)$ are the eigenvalues of the momentum operators $\hat{p}_{\rho}$, i.e., $\hat{p}_{\rho}\psi(x)=p_\rho\psi(x)$. The equation (\ref{free}) implies the relation
\begin{equation}\label{epsilon}
    \epsilon^2={\bf p}^2+m^2.
\end{equation}

To find the normalization factor $N$ we use the definition of the probability density (\ref{density}). Note that due to the choice (\ref{omega}) of the external antisymmetric field $\omega ^{\rho \sigma}(x)$, one finds that
\begin{align}
&b^\rho_\nu(f,g)=ia^\rho\left(-i\partial_\nu f,g\right)+ia^\rho\left( f,-i\partial_\nu g\right)-\partial_\nu a^\rho(f,g)\label{b}\\&=-\frac{i\theta}{4}\eta_{\nu\tau}\varepsilon^{\rho\sigma\tau}( f\partial_\sigma g-\partial_\sigma f g)\notag\\&-\frac{\theta^2}{16}\eta_{\nu\tau}\left(\varepsilon^{\rho\sigma\tau}\varepsilon^{\alpha \beta\upsilon}+\varepsilon^{\rho\sigma\upsilon}\varepsilon^{\alpha \beta\tau}\right)x_\upsilon\left(\partial_\sigma\partial_\alpha f\partial_\beta g-\partial_\alpha f\partial_\sigma\partial_\beta g\right)\notag\\
&-\frac{\theta^2}{48}\left(\eta^{\alpha\rho}\delta^\beta_\nu+\eta^{\beta\rho}\delta^\alpha_\nu\right)\partial_\alpha f\partial_\beta g+O\left( \theta ^{3}\right).\notag
\end{align}

Taking into account the identities
\begin{equation}
    \bar\psi\gamma^\nu\psi=\frac{2p^\nu}{\epsilon+m}N^2,\,\,\,\,\,\bar\psi\psi=\frac{2m}{\epsilon+m}N^2,
\end{equation}
we calculate that
 \begin{equation}
b^0_\nu(\bar\psi\gamma^\nu,\psi)=-N^2\frac{\theta^2}{12} \frac{\epsilon\left(\epsilon^2-{\bf p}^2\right)}{\epsilon+m}+ O\left( \theta ^{3}\right),
\end{equation}
and
\begin{equation}
 \psi^\dagger\star\psi=N^2\frac{2\epsilon}{\epsilon+m}\left[1-\frac{\theta^{2}}{12}\left(\epsilon^2-{\bf p}^2\right)+O\left( \theta ^{3}\right)\right].
\end{equation}
So, the spacial integral of the probability density (\ref{density}) in the state(\ref{s}) is
\begin{equation}
   \int\limits_V\!\varrho(x)\,d^{2}x=N^2V\frac{2\epsilon}{\epsilon+m}\left[1-\frac{\theta^2}{8}\left(\epsilon^2-{\bf p}^2\right)+O\left( \theta ^{3}\right)\right].
\end{equation}
From the normalization condition one finds
\begin{equation}\label{norm}
   N^2V=\frac{\epsilon+m}{2\epsilon}+\frac{\theta^2(\epsilon+m)}{16\epsilon}m^2+O\left( \theta ^{3}\right).
\end{equation}

Now substituting the solution (\ref{s}) in the definition of the energy-momentum tensor (\ref{id7}) one finds for the energy and momentum densities
\begin{eqnarray}
&&T^{00}_\theta=\epsilon N^2\frac{2\epsilon}{\epsilon+m}\left[1-\frac{\theta^2}{8}\left(\epsilon^2-{\bf p}^2\right)+O\left( \theta ^{3}\right)\right], \label{id8}
\\&&T^{0i}_\theta=-p_i N^2\frac{2\epsilon}{\epsilon+m}\left[1-\frac{\theta^2}{8}\left(\epsilon^2-{\bf p}^2\right)+O\left( \theta ^{3}\right)\right]  \notag.
\end{eqnarray}
Integrating these expressions and using the normalization factor $N$ from (\ref{norm}) we end up with
\begin{equation}\label{D3}
  E=\epsilon+O\left( \theta ^{3}\right),\,\,\,\,\,P_i=-p_i+O\left( \theta ^{3}\right).
\end{equation}
Since the eigenvalues of the momentum operators $p_\rho$ satisfy the relation (\ref{epsilon}), we obtain the standard energy momentum dispersion relation
\begin{equation}\label{EMrelation}
    E^2={\bf P}^2+m^2+O\left( \theta ^{3}\right).
\end{equation}
Here we stress that even though the normalized solution (\ref{s}) of the free noncommutative Dirac equation (\ref{free}) is different from the commutative one, the energy-momentum dispersion relation (\ref{EMrelation}) remains valid for the free noncommutative relativistic particle. It happens because the definition of the energy and momenta of the system was also deformed.

Finally we note that the energy-momentum dispersion relation (\ref{EMrelation}) is a consequence of the relativistic invariance of the considered model, i.e., the choice (\ref{omega}) of the antisymmetric field $\omega^{\rho\sigma}\left( x\right)$. If the Poisson structure is not Lorentz covariant, the energy-momentum dispersion relation will change.

\section{Conclusions}

In the present paper we have defined the noncommutative Dirac equation and proved that the basic properties of relativistic wave equations, like the Lorentz covariance and the continuity equation for the probability density remain valid on coordinate dependent noncommutative space-time. Choosing the specific star product and using its properties we have derived the corresponding probability current density and proved its conservation, without using Noether theorem. The energy-momentum tensor for the free noncommutative spinor field was calculated. The example of the free noncommutative fermionic field show that in the first orders in $\theta$ the standard energy-momentum dispersion relation $E^2={\bf P}^2+m^2$ is recovered.

In conclusion we would like to address some questions concerning the physical meaning of the proposed model. What is the form of the gauge transformation leaving the electromagnetic coupling invariant and how to describe the dynamics of the corresponding gauge field. The standard abelian $U(1)$ gauge symmetry is broken in the canonical noncommutative field theory. To restore the gauge invariance one should substitute $U(1)$ with the non-abelian group $U_\star (1)$. If the noncommutativity parameter is point dependent the situation becomes much more complicated. In \cite{Wess} it was shown that to make the field theory on $\kappa$-Minkowski space-time gauge invariant one should admit the derivative dependence of the gauge field $A_\rho$. The proper gauge transformation also becomes derivative valued. To obtain this result the authors of \cite{Wess} used the explicit form of the co-product for the $\kappa$-Minkowski space-time. In general case of the Kontsevich star product the co-product is unknown. One may use the perturbative techniques \cite{kup14,kup15} to construct the gauge transformation leaving invariant the noncommutative field theory of the general form. To study the dynamics of the corresponding gauge field one needs to employ the spectral action techniques.

Another important problem is the nature of the antisymmetric field $\omega^{\rho\sigma}\left( x\right)$ and its connection with the gravity $g^{\rho\sigma}\left( x\right)$. Here we mention some interesting ideas in this respect. The authors of \cite{Freidel} have shown that the effective dynamics of matter fields coupled to $3d$ quantum gravity after integration over the gravitational degrees of freedom is described by a braided noncommutative quantum field theory. In \cite{Steinacker} the noncommutativity links the compact and noncompact dimensions providing the mechanism for emergent gravity on the brane solutions in Yang-Mills matrix models. In \cite{Vasilevich} the holographic duality was descovered between the Poisson sigma models, which are equivalent in the specific configuration of the target space to the two-dimensional dilaton gravities, and the quantum mechanics with coordinate dependent noncommutativity.

\section*{Acknowledgements}

I am grateful to Dima Vassilevich for useful comments. This work was supported in part by FAPESP.

\end{document}